\documentstyle[twocolumn,prl,epsf,aps]{revtex}

\newcommand{\rmd}{{\rm d}}
\newcommand{\rmi}{{\rm i}}

\newcommand{\beq}{\begin{equation}}
\newcommand{\eeq}{\end{equation}}
\newcommand{\bea}{\begin{eqnarray}}
\newcommand{\eea}{\end{eqnarray}}

\newcommand{\cuoo}{CuO$_{2}$}

\newcommand{\lsco}{La$_{2-x}$Sr$_{x}$CuO$_{4}$}
\newcommand{\ybco}{YBa$_{2}$Cu$_{3}$O$_{6+x}$}

\begin{document}                                                

\wideabs{

\draft

\title{Charge pairing, superconducting transition and supersymmetry
in high-temperature cuprate superconductors}

\author{Eduardo C. Marino and Marcello B. Silva Neto}

\bigskip

\address{
Instituto de F\protect\'\i sica, Universidade Federal do Rio de Janeiro,
Caixa Postal 68528, Rio de Janeiro - RJ, 21945-970, Brazil}

\date{\today}
\maketitle


\begin{abstract} 

We propose a model for high-T$_{c}$ superconductors, 
valid for $0\leq\delta\leq\delta_{SC}$, that includes both the
spin fluctuations of the Cu$^{++}$ magnetic ions and of the O$^{--}$ doped holes. 
Spin-charge separation is taken into 
account with the charge of the doped holes being associated to quantum skyrmion 
excitations (holons) of the Cu$^{++}$ spin background. The holon effective 
interaction potential is evaluated as a function of doping, indicating that
Cooper pair formation is determined by the competition between the spin 
fluctuations of the Cu$^{++}$ background and of spins of the O$^{--}$ doped 
holes (spinons). The superconducting transition occurs when the spinon fluctuations 
dominate, thereby reversing the sign of the interaction. At 
this point ($\delta = \delta_{SC}$), 
the theory is supersymmetric at short distances and, as a consequence, 
the leading order results are not modified by radiative corrections. The 
critical doping parameter for the onset of superconductivity at $T=0$ is obtained
and found to be a universal constant determined by the shape of the Fermi surface. Our 
theoretical values for $\delta_{SC}$ are in good agreement with the experiment 
for both LSCO and YBCO.

\end{abstract}


}               

\begin{narrowtext}


{\it Introduction.}
High-temperature superconducting cuprates have a very rich and complex phase 
diagram whose understanding is an important issue. 
In the underdoped region, for instance, a wide variety of physical 
phenomena like N\'eel and metal-insulator transitions, transport (non-Fermi 
liquid) anomalies, the occurrence of a spin-pseudogap, absence of a sharp 
quasiparticle peak (spin-charge separation), etc., have inspired a large 
amount of theoretical and experimental work for about fifteen years 
\cite{Review-I}. In spite of that, even the nature of the 
ground state and of its elementary excitations has not yet been fully determined 
and many different pictures are availabe, ranging from a dimerized ground 
state with spin-Peierls or valence-bond order \cite{Read-Sachdev} 
until the so called staggered-flux ($d$-wave) phase \cite{SF-Phase}. 

Another fundamental point yet to be understood is the mechanism of 
charge pairing leading to the superconducting transition. In connection
to this, we stress that it is by now well established that antiferromagnetic 
spin correlations play an important r{\^ o}le in the dynamics of the system, 
even after the destruction of the N\'eel state. Indeed, different spin-fluctuation 
models have been successfully used to explain the observed spectral weight in 
ARPES data in the quantum disordered phase of high-T$_{c}$ materials 
\cite{Spectral-Weight}, as well as other anomalies \cite{Anomalies}. Moreover, 
the idea of spin-fluctuation induced charge pairing and superconductivity has been 
used recurrently \cite{Spin-Fluctuation}. 

In this work we propose a model for high-T$_{c}$ cuprates valid for 
dopant concentrations ranging from zero up to the superconducting transition, 
$0\leq\delta\leq\delta_{SC}$, that takes into account the spin fluctuations 
of the Cu$^{++}$ magnetic ions and of the O$^{--}$ doped holes on different 
footing, as suggested by the different temperature dependences of the NMR 
relaxation rates for the Oxygen and Copper spins \cite{NMR}. Our model 
also incorporates spin-charge separation \cite{Spin-Charge} as follows. 
The charge of the dopants introduced in the {\cuoo} planes  
is associated to skyrmion quantum spin excitations of the Cu$^{++}$ background (holons)
which, in the N\'eel phase appear as finite energy defects closely related to 
their classic counterparts, whereas in the quantum disordered phase are nontrivial zero 
energy purely quantum mechanical excitations. The spin of the doped holes, 
on the other hand, is represented by chargeless, massless Dirac fermion fields (spinons)
\cite{Marston-Affleck-Lee-Nagaosa-Kim}. We then calculate 
the effective interaction potential between 
the quantum skyrmion topological excitations, as a function of doping,
in order to study charge pairing. 
It becomes clear that Cooper pair formation at zero temperature is controlled by 
the competition between two different contributions to the quantum skyrmion 
effective interaction energy; one coming from the spin fluctuations of Cu$^{++}$ 
magnetic ions and the other from the corresponding fluctuations of the spins of 
the doped holes (spinons). The superconducting transition occurs when the latter 
dominates, thereby reversing the sign of the effective quantum skyrmion (dopant 
charge) interaction potential, at short distances, from {\it repulsive} to 
{\it attractive}. Interestingly, the model becomes supersymmetric at short 
distances, precisely at the onset of superconductivity, thereby making our 
one-loop results robust against radiative corrections, in accordance to 
Witten's theorem \cite{Witten}. The critical value of the doping parameter for 
the superconducting transition at zero temperature, $\delta_{SC}$, is universally 
determined by the shape of the Fermi surface and is in good agreement with 
experiment for both {\lsco} and {\ybco} compounds. The model also correctly 
predicts the zero temperature 
magnetization curves as a function of the critical doping in the AF ordered phase.

{\it The model.}
In previous works \cite{Marino-Marcello}, we have proposed a model for doping quantum 
Heisenberg antiferromagnets that successfully described the magnetization curves and 
the AF part of the phase diagrams of the two best studied high-T$_{c}$ compounds, 
namely LSCO and YBCO. One of the important consequences of that model is the
observation that each hole added to the {\cuoo} planes 
creates a skyrmion topological defect, as has 
been proposed earlier \cite{Skyrmion-Literature}. The dopant charge, in particular, is 
attached to the skyrmion charge and its dynamics becomes totally determined by the 
quantum skyrmion correlation functions. In the N\'eel ordered Mott insulating 
phase, the skyrmions have a finite excitation energy and this reflects the 
existence of a gap for charge that can be associated to the
breakdown of translational invariance of the lattice. The model proposed in 
\cite{Marino-Marcello}, however, is restricted to the antiferromagnetic part 
of the phase diagram, where $\delta\leq\delta_{AF}$. Nevertheless, we shall 
pursue the picture in which skyrmions are in general the charge carriers of 
the doped holes. In particular, we shall exploit this idea in the quantum disordered 
phase, $\delta_{AF}\leq\delta\leq\delta_{SC}$, where the skyrmions are purely 
quantum mechanical and have zero energy. This is again consistent with 
charge response experiments, like in-plane optical conductivity \cite{optcond}, 
which indicates the absence of a gap for charge excitations in the quantum 
disordered metallic phase. 

Let us consider the zero temperature
Euclidean partition function 
\beq
{\cal Z}=\int{\cal D}\bar{z}{\cal D}z{\cal D}\overline{\psi}{\cal D}\psi
{\cal D}{\cal A}_{\mu}{\;}\delta[\bar{z}z-1]
e^{-S(\bar{z},z,\overline{\psi},\psi,{\cal A}_{\mu})},
\label{Part-Func}
\eeq
where 
\bea
S(\bar{z},z,\overline{\psi},\psi,{\cal A}_{\mu})&=&\int\rmd^{3}x
\left\{\frac{1}{g_{0}} \left|(\partial_{\mu}-\rmi{\cal A}_{\mu})z_i\right|^{2}
\right. \nonumber \\ &+& \left. 
\overline{\psi}_{a} {\;} \gamma_{\mu} \left(\rmi\partial^\mu + q
{\cal A}^{\mu}\right)\psi_{a} \right\},
\label{Effective-Action}
\eea
and we use units in which $\hbar=c_{0}=1$. In the above expression $z_i^{\dag},z_i$, 
$i =1,2$, are Schwinger boson fields related to the local spin density of Cu$^{++}$
ions through ${\bf S}=z_{i}^{\dag}${\mathversion{bold}$\sigma$}$_{ij} z_{j}$, 
and $\psi^{\dag}_{a},\psi_{a}$, 
$a=\uparrow,\downarrow$, are chargeless $2$-component 
Dirac spinor fields (spinons) describing the local 
spin density of the doped O$^{--}$ holes through 
${\bf S}_h=\psi_{a}^{\dag}${\mathversion{bold}$\sigma$}$_{ab}\psi_{b}$. 
As usual, ${\cal A}_\mu$ is the Hubbard-Stratonovitch field and $g_{0}$ the bare coupling 
constant of the CP$^{1}$ model. The constant $q$ measures the strength of the 
coupling of spinons.

Spin-charge separation is manifested in our model through the fact that the
massless Dirac fermions carry the spin of the doped holes, whereas
all information about their charge is carried 
by the {\it quantum }skyrmion excitations (holons) created out of the
Schwinger boson background \cite{Marino-Marcello}.
The description of spinons as massless Dirac fields arises
naturally in the continuous limit of microscopic models 
\cite{Marston-Affleck-Lee-Nagaosa-Kim}. The full treatment of 
the quantum skyrmions of the theory described by (\ref{Effective-Action}), 
on the other hand, has been carried out in \cite{Marino}.

Before we proceed, it is important to determine how the doping dependence will be 
introduced in our theory. As we explained above, we may identify in principle, at 
least two phases in the model given by (\ref{Effective-Action}), at $T=0$. An ordered 
N\'eel phase, $g_{0}<g_{c}$, for which there is a nonzero spin stiffness 
$\rho_{s}=1/g_{0}-1/g_{c}>0$ ($g_{c}$ is the quantum critical coupling) and 
a quantum disordered phase, $g_{0}>g_{c}$, in which the Schwinger bosons 
$z_i$ acquire a mass $m^2 \propto [1/g_{c}-1/g_{0}]$, and $\rho_{s}=0$. 
As we explained in \cite{Marino-Marcello}, the whole dynamics of the in-plane 
dopant charge is identical the quantum dynamics of skyrmions. 
The quantum skyrmion correlation function corresponding to (\ref{Effective-Action}) 
has been evaluated in \cite{Marino}, in the ordered phase, giving
\beq
\langle\mu(x)\mu^{\dag}(y)\rangle=\frac{e^{-2\pi\rho_{s}|x-y|}}{|x-y|^{q^2/4}},
\label{sk-cf1}
\eeq
where $\mu^{\dag}$ is the quantum skyrmion creation operator.
Conversely, for the theory studied in \cite{Marino-Marcello} the corresponding correlator
was found to be
\beq
\langle\mu(x)\mu^{\dag}(y)\rangle=
\frac{e^{-2\pi\rho_{s}(\delta)|x-y|}}{|x-y|^{\alpha(\delta)}},
\label{sk-cf}
\eeq
where the expressions for $\rho_{s}(\delta)$ and $\alpha(\delta)$ 
have been determined in \cite{Marino-Marcello}. In particular,
\beq
\alpha(\delta) = 
\left[\frac{64}{\pi^2+16}+\frac{\alpha_{EM}}{4\pi^2}\right](n \delta)^2
\label{alfa}
\eeq
with $n = 1$ for YBCO and $n = 4$ for  LSCO, the factor of four being a 
consequence of the existence of four branches in the Fermi surface for 
this compound, see discussion in \cite{Marino-Marcello}. The $\rho_{s}(\delta)$ 
function is given by $\rho(\delta)=\rho(0)[1 - A\delta^2]$, for YBCO and 
$\rho(\delta)=\rho(0)[1 - B\delta - C\delta^2]^{1/2}$, for LSCO, and again the 
different behavior being ascribed to the form of the Fermi surface in each 
case \cite{Marino-Marcello}. The constants $A$,$B$ and $C$ have been 
evaluated from first principles in \cite{Marino-Marcello}. In 
(\ref{alfa}), $\alpha_{EM}$ is the electromagnetic fine structure constant 
and accounts for the contribution of the electromagetic interaction of 
the doped holes to the skyrmion correlation function. Examining (\ref{alfa}) 
we see that actually this term can be neglected when compared to the first 
one. In order to obtain the $\delta$-dependence of the spin stiffness 
$\rho_{s}$ and of the spinon coupling $q$ in our model (\ref{Effective-Action}), 
we now match the two correlation functions in (\ref{sk-cf1}) and 
(\ref{sk-cf}) (ordered phase), obtaining
$\rho_s=\rho(\delta)$ and $q=[\frac{256}{\pi^2+16}]^{1/2}(n\delta)$, 
where we have already neglected the electromagnetic part. We immediately 
conclude that the sublattice magnetization in the ordered phase is given 
by $M(\delta)=\sqrt{\rho(\delta)}$. From this we can readily obtain $\delta_{AF}$ 
from $\rho(\delta_{AF})=0$, see also \cite{Marino-Marcello}. For
$\delta >\delta_{AF}$, on the other hand, 
where $\rho_s = 0$, we assume that the expression for $q(\delta)$
still holds. 

{\it Cooper pair formation.} 
Let us now investigate the conditions for skyrmion pairing 
and consequent formation of Cooper pairs, by analyzing the effective interaction 
potential between quantum skyrmions in the quantum disordered 
underdoped phase. For this purpose, we introduce the skyrmion 
current ${\cal J}^{\mu} = \frac{1}{2\pi}\epsilon^{\mu\alpha\beta}
\partial_{\alpha}{\cal A}_{\beta} $ through the identity
\bea
{\cal Z}&=&\int{\cal D}{\cal J}_{\mu}{\cal D}{\cal A}_{\mu}
{\cal D}\bar{z}{\cal D}z
{\cal D}\overline{\psi}{\cal D}\psi{\;}
\delta[{\cal J}_{\mu}-\frac{1}{2\pi}\epsilon^{\mu\alpha\beta}
\partial_{\alpha}{\cal A}_{\beta}] \nonumber \\
&\times& e^{-S[{\bar{z},z,\overline{\psi},\psi,\cal A}_{\mu}]}.
\label{zja}
\eea
Integrating over $z_i^{\dag},z_i$ and 
$\overline{\psi}_a,\psi_a$, we obtain, at one-loop level, the effective 
action
\beq
S_{eff}[{\cal A}_{\mu}]=\int \rmd^{3}x\int\rmd^{3}y{\;}
\left\{\frac{1}{4}{\cal F}_{\mu\nu}(x)\Sigma(x-y)
{\cal F}_{\mu\nu}(y)\right\},
\eeq
where the kernel $\Sigma(x-y)$ has Fourier transform
given by $\Sigma(p)=\Pi_{B}(p)+\Pi_{F}(p)$ where
\beq
\Pi_{B}(p)=\frac{1}{2\pi}\left[\frac{m}{p^{2}}-
\frac{1}{2p}\mbox{arctan}\left(\frac{p}{m}\right)-
\frac{2m^{2}}{p^{3}}\mbox{arctan}\left(\frac{p}{2m}\right)\right],
\label{pib}
\eeq
and
\beq
\Pi_{F}(p)=\frac{1}{2\pi}\left[\frac{q^{2}}{8p}\right].
\label{pif}
\eeq
These two terms are, respectively, the 
contributions to the vacuum polarization coming from 
the complex scalar fields $z_i$ (Schwinger bosons) and
fermions $\psi$ (spinons). In (\ref{pib}), $m$ is the mass of the 
$z_i$-fields (spin-gap) in the quantum disordered phase, where
$\delta>\delta_{AF}$ ($g_0>g_{c}$)

In order to obtain the effective skyrmion action, we use an 
exponential representation for the $\delta$-function in (\ref{zja})
and integrate over the corresponding Lagrange multiplier field 
and ${\cal A}_\mu$. The result is
\beq
{\cal Z}=\int{\cal D}{\cal J}_{\mu}{\;}e^{\left\{-\frac{(2\pi)^{2}}{2}
\int\rmd^{3}x\int\rmd^{3}y{\;}{\cal J}_{\mu}(x){\Sigma}^{\mu\nu}(x-y)
{\cal J}_{\nu}(y)\right\}},
\label{Current-Current}
\eeq
where the Fourier transform of the kernel is given by
\beq
{\Sigma}^{\mu\nu}(p)=
\frac{\Sigma(p)}{p^{2}}\left (p^{2}\delta^{\mu\nu}-p^{\mu}p^{\nu}\right ).
\eeq

From expression (\ref{Current-Current}) we can 
readily obtain the effective interaction 
energy between static skyrmions. This is given, in real time, by
\beq
{\cal H}_{I}=-\frac{1}{2}(2\pi)^{2}\int\rmd^{2}{\bf x}\int\rmd^{2}{\bf y}
\rho({\bf x})K({\bf x}-{\bf y})\rho({\bf y}),
\eeq
where $\rho({\bf x}) = {\cal J}_{0}({\bf x},0)$ is the dopant charge density and 

$$K({\bf x}-{\bf y})=\int\rmd^{2}{\bf p}/(2\pi)^{2}{\;}
e^{\rmi {\bf p}\cdot({\bf x}-{\bf y})}{\;}\Sigma({\bf p},0).
$$

For two charges at positions ${\bf x}_{1}$ 
and ${\bf x}_{2}$, we have $\rho({\bf x})=\delta^{(2)}
({\bf x}-{\bf x}_{1})+\delta^{(2)}({\bf x}-{\bf x}_{2})$. After
discarding self-interactions, we obtain the effective interaction
potential for static charges, namely
\beq
V({\bf x}_{1}-{\bf x}_{2}) = 
\int \rmd^{2}{\bf p}{\;}
\Sigma({\bf p},0){\;}e^{\rmi{\bf p}\cdot({\bf x_1} - {\bf x_1})},
\label{Int-Potential}
\eeq
where
\bea
\Sigma({\bf p},0)&=& - \frac{q^{2}}{8|{\bf p}|}+
\frac{1}{2\pi}\left[
\frac{1}{2|{\bf p}|}\mbox{arctan}\left(\frac{|{\bf p}|}{m}\right)
- \frac{m}{|{\bf p}|^{2}}
\right. \nonumber \\ &+& \left.
\frac{2m^{2}}{|{\bf p}|^{3}}\mbox{arctan}\left(\frac{|{\bf p}|}
{2m}\right)\right].
\eea

It is well known that in high-T$_c$ superconducting cuprates, Cooper pairs 
form in such a way that the two charges are localized in space. Only the short
distance behavior of the interaction potential, therefore, is relevant for
Cooper pair formation. In this limit (large $|{\bf p}|$) we have
\beq
V({\bf x}_{1}-{\bf x}_{2})\rightarrow \int\rmd^{2}{\bf p}
\left[\frac{1}{8|{\bf p}|} - \frac{q^{2}}{8|{\bf p}|}\right]
{\;}e^{\rmi{\bf p}\cdot ({\bf x}_{1}-{\bf x}_{2})}.
\label{Effective-Potential}
\eeq
The first contribution inside the square brackets in the above expression corresponds 
to the Cu$^{++}$ spin fluctuations (Schwinger bosons) while the second corresponds 
to fluctuations from O$^{--}$ spins (spinons). We see that for small doping, 
$q^{2}<1$, the potential is allways {\it repulsive} and 
there is no charge pairing. For $q^{2}>1$, on the other hand, the interaction 
potential becomes {\it attractive} and charge (skyrmion) pairing occurs. 
Consequently, we conclude that the critical doping for the onset of 
superconductivity is determined by the condition  
\beq
q^{2}(\delta_{SC})=1.
\label{dsc}
\eeq
Let us remark that if we had considered a system of spinons solely, without including 
the Cu$^{++}$ background, we would have obtained that the interaction potential 
(\ref{Effective-Potential}) would
always be attractive for any $q\neq 0$, at zero temperature, and $\delta_{SC}=0$. 
This is what happens in the mean field phase diagram of Kotliar and Liu for the 
$t-J$ model \cite{Kotliar-Liu}. We see that the primer effect of considering 
the Cu$^{++}$ spin background is to shift the value of $\delta_{SC}$ to the right 
of the phase diagram, which is actually what is observed experimentally.

{\it Comparison with experiment.}
From the expression of $q$ in terms of $\delta$ ($q = 2 \sqrt{\alpha(\delta)}$, 
see (\ref{alfa})), we may infer that $\delta_{SC}$ is an universal constant, 
only determined by the shape of the Fermi surface. We see, in particular, that 
$\delta_{SC}^{YBCO}=4\delta_{SC}^{LSCO}$, a relation that is 
verified by experiments, if we take in account the relation between $\delta$ and 
the stoichiometric doping parameter $x$, namely $\delta=x$ for LSCO and 
$\delta=x-0.20$ for YBCO. Another prediction of our model is that compounds 
with similar Fermi surfaces should have the same superconducting critical 
doping $\delta_{SC}$.
From (\ref{dsc}) and (\ref{alfa}), we calculate $\delta_{SC}^{YBCO}=0.318$ and 
$\delta_{SC}^{LSCO}=0.079$, which have a fairly good agreement with experiment. 
We show below that taking in account the presence of disorder, we can obtain 
better values for these critical doping parameters.

{\it Disorder.} Disorder may be modelled in the ordered N\'eel phase of 
a doped antiferromagnet by considering a continuous random distribution of spin
stiffnesses \cite{em}. If we introduce a Gaussian$\times |\rho - \rho_s|^{\nu -1}$ 
distribution, with exponentially suppressed magnetic dilution, in the original 
model \cite{Marino-Marcello} used to describe the antiferromagnetic phase, we obtain 
a correction for (\ref{sk-cf}), namely 
$\alpha(\delta)\rightarrow \alpha'(\delta) = \alpha(\delta) + \nu$ \cite{em}. 
Choosing $\nu = \frac{1}{8}$ for both compounds, we get 
\beq
\delta_{SC} = \frac{1}{n} \sqrt{ \frac{\pi^2 +16}{512} }
\eeq
Now, the critical doping parameters at $T=0$ become
$\delta_{SC}^{YBCO}=0.225$ and $\delta_{SC}^{LSCO} = 0.056$, corresponding to
$x_{SC}^{YBCO} = 0.425$ and $x_{SC}^{LSCO} = 0.056$, which are in good agreement
with experiment. Notice that with a single choice for the disorder distribution
we correctly obtained the critical dopings for both YBCO and LSCO.

{\it Supersymmetry.} Let us observe now a remarkable fact. At short distances, when the
mass of the Schwinger bosons, $m$, may be neglected, our model becomes supersymmetric
precisely at the point $q(\delta_{SC})=1$, where the superconducting transition occurs. 
Supersymmetry relates Schwinger bosons $z_i$ and spinons $\psi_{a}$ and that is why the 
contributions of both to the holon (skyrmion) interaction potential
are identical but with opposite signs at this point. An important consequence 
is that our one-loop derivation of the holon effective interaction 
potential and critical dopings
are unchanged by radiative corrections. Indeed, we have actually checked that the 
contribution of these corrections to the short distance behavior of the effective 
holon interaction potential (\ref{Effective-Potential}) is 
subdominant and can be neglected. This can be 
understood on general grounds, as a result of Witten's theorem \cite{Witten}, which
states that supersymmetry cannot be broken perturbatively.

{\it Conclusions.} We have calculated the effective interaction potential between
holons in a spin-charge separated, spin fluctuation model for hith-T$_{c}$ cuprates.
We have shown that Cooper pair formation and superconductivity is determined by the 
competition between the spin fluctuations of the Cu$^{++}$ antiferromagnetic background 
and the spin fluctuations of the doped O$^{--}$ holes, in the underdoped regime. 
Our prediction of the critical doping for the onset of superconductivity at zero
temperature, $\delta_{SC}$, is in good agreement with experiment for 
either LSCO and YBCO compounds. We stress that the pairing must be between skyrmions
and not between skyrmion and anti-skyrmion, so that the total electric charge of 
the pair is nonzero. 

At finite temperatures, the pairing shall no longer occur for $q^{2}(\delta)=1$. 
There will be finite temperature corrections for both $\Pi_{B}$ and $\Pi_{F}$ and, since
supersymmetry is broken at any finite temperature, the fermionic and bosonic contributions 
for the interaction potential should cancel at $q \neq 1$ ($\delta \neq \delta_{SC}$), 
in agreement with experimental results. We are presently investigating
this point.

\end{narrowtext}


This work has been supported in part by  FAPERJ and 
PRONEX - 66.2002/1998-9. E. C. M. was partially supported by CNPq
and M. B. S. N. by FAPERJ.


\end{document}